\documentstyle[prl,aps,twocolumn,psfig]{revtex}

\begin{document}

\draft
\preprint{submitted to Phys.Rev.Lett.}

\twocolumn[\hsize\textwidth\columnwidth\hsize\csname @twocolumnfalse\endcsname
\title{ 
Averaging theory for the 
structure of hydraulic jumps and separation\\
in laminar free-surface flows}

\author{Tomas Bohr, Vachtang Putkaradze and Shinya Watanabe}
\address{Center for Chaos \& Turbulence Studies,
Niels Bohr Institute, Blegdamsvej 17, Copenhagen, 2100, Denmark}

\date{to appear in Phys.Rev.Lett., vol.79, 1038 (1997 Aug.\ 11)}
\maketitle

\begin{abstract}
We present a simple viscous theory 
of free-surface flows in boundary layers,
which can accommodate regions of separated flow. In particular this 
yields the structure of stationary hydraulic jumps, both in their
circular and linear versions, as well as structures moving with
a constant speed. 
Finally we show how the fundamental hydraulic concepts
of subcritical and supercritical flow, 
originating from inviscid theory,
emerge at intermediate length scales in our model.
\end{abstract}
\vskip2pc]

\narrowtext 

Despite the classical nature of the subject, the flow of a viscous fluid
with a free surface presents many unsolved theoretical
problems, even under laminar conditions. To a large extent this is 
due to the lack of approximate methods for describing flows 
containing {\em separated} regions, 
i.e.\ regions in which the flow is reversed 
with respect to the mean flow. 
Hydraulic jumps are examples of such flows. 
They are large, sudden deformations 
in the free surface of stationary
flows \cite{c1:riverbore} and no theory exists for their 
structure --- save the full Navier-Stokes equations
combined with the free-surface boundary conditions, 
for which even numerical solution poses large problems.
In this Letter we present a method for determining
some of these flows,
which includes viscosity and variations of the velocity profile.
Stationary states are obtained as trajectories
in a simple two-dimensional phase space.

An inviscid theory of hydraulic jumps, which is still the standard 
hydraulic approach to the subject, is due to Lord Rayleigh in 1914 
\cite{R1914,hydref}. He regarded hydraulic jumps as discontinuities 
(shocks) which can occur in the shallow water equations
\cite{Whitham}.
Across a jump the flow decelerates 
from a rapid {\em supercritical} flow, 
in which disturbances propagate only down stream, 
to a {\em subcritical} flow, 
in which they propagate in both directions.

The {\em circular hydraulic jump} is 
easy to study 
experimentally and to maintain in a laminar state. 
Here a jet of fluid falls vertically 
onto a horizontal surface (Fig.~1). 
The fluid spreads in an axisymmetric way, 
and a hydraulic jump is formed at some $r_j$. 
The value of $r_j$ cannot be found by the standard theory, 
and it depends strongly on viscosity $\nu$ \cite{BDP}. 
Further, experiments show clearly that a 
{\em separation bubble}, or a recirculating region,
forms on the bottom in conjunction with the jump 
\cite{BDP,Tani,CraikBowles,Ellegaard}. 

Separation {\em per se} has been studied more intensely 
in {\em boundary layers} close to solid bodies, 
e.g.\ airfoils,
immersed in a high Reynolds number flow.
This line of research descends from Prandtl's seminal work 
\cite{Schlichting}
in which he introduced 
the boundary layer approximation --- 
a radical simplification of the Navier-Stokes equations.
Prandtl's equations are
valid in a thin layer near solid (no-slip) surfaces, 
where the fluid motion is predominantly along the surface.
But in general, they become {\em singular} at 
separation points \cite{singularities}, 
where the assumption of forward flow breaks down.
The boundary layer equations can be further simplified
by using an averaging technique of von Karman and Pohlhausen 
\cite{Schlichting}, 
in which the tangential velocity profile is 
approximated as a low order polynomium. 
This model is useful up to
a separation point,
but solutions beyond the point tend to diverge and are discarded.
Such troubles can be cured by
taking into account the feed-back effect from the 
boundary layer on the external flow.
The ``inverse method'' \cite{inverse} makes it possible to
calculate flows with separation bubbles,
and analytical results have been obtained for the structure
of separation points at large Reynolds numbers \cite{tripledeck}.

It is natural to employ the boundary layer approximation
to describe hydraulic jumps
since the fluid moves nearly parallel to
the bottom surface.
To avoid the singularities near a jump encountered in earlier work
\cite{Tani,BDP} we use the Karman-Pohlhausen method.
The standard hydrostatic approximation then provides the link
between pressure and layer thickness
analogous to the feedback mechanism in
the inverse method used above \cite{long}.

For a stationary, radially symmetric flow with a free surface the
boundary layer equations take the form
\begin{equation}
u u_r + w u_z = -g h' + \nu u_{zz}
\label{momentum}
\end{equation}
\begin{equation}
\label{continuity}
u_r + u/r + w_z = 0
\end{equation}
where $u(r,z)$ and $w(r,z)$ are the radial ($r$) 
and vertical ($z$) velocity components, 
$h(r)$ is the height, 
and we assume hydrostatic pressure.
Surface tension has been neglected, since it does not appear 
to be decisive in determining the structure of the flow,
although it is 
necessary for the stability of the flows as discussed below.
The boundary conditions are no-slip on the bottom: 
$u(r,0) = w(r,0) = 0$,
no stress on the top: $u_z(r,h) = 0$ 
(strictly valid 
only for  small deformations $|h'|$) 
and the kinematic boundary condition 
at the top:
$w(r,h) = u(r,h) h'$, which ensures the mass conservation:
$2 \pi r \int_0^h u dz = $const.$ = Q = 2 \pi q$.
By rescaling the horizontal and vertical lengths and 
velocities by
$L = (q^5 \nu^{-3} g^{-1})^{1/8}$,
$H = (q \nu g^{-1})^{1/4}$,
and $V = (q \nu g^3)^{1/8}$, respectively,
all parameters are eliminated 
from (\ref{momentum},\ref{continuity}) \cite{BDP}.

We average these equations over $z$, but in 
contrast to earlier approaches \cite{Tani,BDP} we shall not assume 
a self-similar velocity profile. 
Instead the velocity profile is parametrized as
\begin{equation}
u(r,z) = v(r) \left( a_1(r) \eta + a_2(r) \eta^2 + a_3(r) \eta^3 \right)
\end{equation}
where $\eta = z/h(r)$. 
Implementing the boundary conditions reduces the 
parameters to just one:  $\lambda (r)$, 
i.e.\ $a_1=\lambda + 3$, $a_2=-(5 \lambda + 3)/2$ and
$a_3=4 \lambda/3$.
Thus the profile is parabolic when $\lambda = 0$ and
separation occurs for $\lambda = -3$. 
With these assumptions the averaged momentum equation 
(\ref{momentum}) takes the form \cite{long}
\begin{equation}
\label{lambda1}
  v(F_2(\lambda) v)' = -h' - (\lambda+3)v/h^2
\end{equation}
where 
$F_2(\lambda) = (\int_0^h u^2 dz)/(h v^2) = 6/5 - \lambda/15 + 
\lambda^2/105$
and $rhv = 1$. 
To determine $\lambda$, one more equation is needed, 
and this is 
(as in the standard Karman-Pohlhausen approach \cite{Schlichting}) 
taken to be (\ref{momentum}) evaluated on the bottom ($z=0$), 
which gives
\begin{equation}
\label{lambda2}
  h' = -(5 \lambda + 3)v/h^2 .
\end{equation}
When $v=1/(rh)$ is inserted into (\ref{lambda1},\ref{lambda2}) we 
obtain a non-autonomous flow for the two-dimensional vector field 
$(h,\lambda)$. The flow has singularities only on the lines $h=0$ 
and 
$\lambda = 7/2$.  
It is thus possible to obtain both separation and 
parabolic profiles without crossing singularities.

\begin{figure}[tbp]
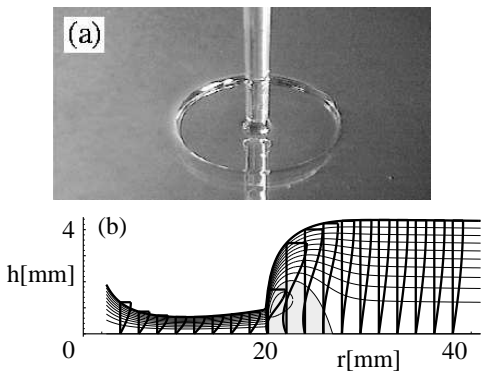

\centerline{\psfig{file=fig1a.epsi,height=1.1in}}
\centerline{\psfig{file=fig1b.epsi,height=0.85in}}
\vspace{3mm}
\caption{
(a) A circular hydraulic jump is formed
when a liquid jet falls onto a plate from above
(photo: courtesy of A.\ E.\ Hansen).
(b) Surface profile, streamlines,
and the horizontal velocity profile in a cross section,
predicted from our model (see text and 
Fig.~\protect{\ref{radialresult1}}).
Note the difference in the scales for the two axes.
The profile is nearly parabolic at large radius,
but is strongly deformed near the jump.
The shaded area is a separation bubble.}
\label{pattern}
\end{figure}

\begin{figure}[tbp]
\centerline{\psfig{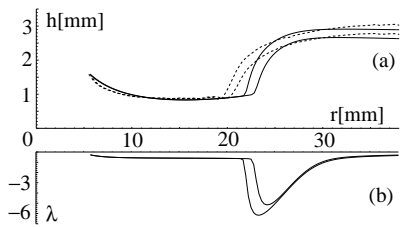}}
\vspace{3mm}
\caption{
(a) Measured $h(r)$ (dashed curves)
vs.\ our model (solid curves).
The model uses a shooting method from $r_2=30$[mm]
toward $r_1=12$[mm] for a fixed inner height $h_1$ 
and two outer heights $h_2$.
Fluid: 50\% ethylene glycol ($Q=27$[m$\ell$/s], 
$\nu=7.6\times 10^{-6}$[m$^2$/s]).
Length and velocity scales: $L=28$[mm], $H=1.4$[mm], and 
$V=12$[cm/s].
(b) The model predicts a larger separation zone
as $h_2$ increases.
}
\label{radialresult1}
\end{figure}

We solve the system with two boundary conditions.
Since the velocity profiles are not measured,
we impose two surface points $h_1(r_1)$ and $h_2(r_2)$,
read from a recent measurement of
Ellegaard {\it et al.}~\cite{Ellegaard}.
Iterative adjustment of $\lambda$ at one end 
converges to a solution
which passes through the two chosen points.
Figure \ref{radialresult1}(a) shows comparison
of the calculated height $h(r)$
and the measurements, for two different $h_2$ values.
The surface profiles near $r_j$ show fair agreements,
considering the simplicity of the model,
but $r_j$ is off by around 15\%.
Figure \ref{radialresult1}(b) shows
the calculated $\lambda(r)$.
A separation zone ($\lambda<-3$) occurs just behind the jump 
and its size increases as $h_2$ is raised,
just as observed.
The streamlines and velocity profiles are determined
from $\lambda$, and this leads to a graphical representation
of the flow as shown in Fig.~\ref{pattern}(b).

The model captures the experimental feature
that $h(r)$ inside the jump is little affected 
by the change in $h_2$.
The curves in Fig.~\ref{radialresult1}(a) apparently follow a 
single curve inside the jump,
from which trajectories diverge when $r$ is increased.
Backward integration from $r_2$
automatically settles down to this value of $\lambda$,
which helps us since the entrance velocity profile 
need not be specified.
Further details of the phase space structure 
can be found in \cite{long}.

\begin{figure}[tbp]
\centerline{\psfig{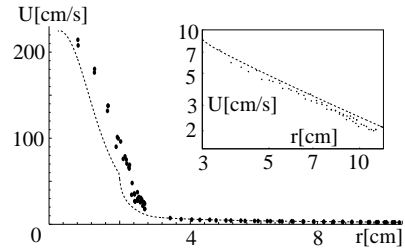}}
\vspace{3mm}
\caption{
Measured surface velocity $U(r)$ (dots)
vs.\ our model (dotted lines).
Parameters: 80\% ethylene glycol ($Q=34$[m$\ell$/s],
$\nu=14.4\times 10^{-6}$[m$^2$/s]), 
$L=25$[mm], $H=1.7$[mm], and $V=16$[cm/s].
The jump is located at $r_j \approx 24$[mm] (experiment)
and 20[mm] (model).
The inset is an enlargement at large $r$ in a log--log scale.
}
\label{radialresult2}
\end{figure}

Another quantity measured in \cite{Ellegaard}
is the surface velocity $U(r) = u(r,h(r))$.
Comparison is made in Fig.~\ref{radialresult2},
which shows quantitative agreement,
although $r_j$ again comes out smaller.
There is no free parameter other than $h_{1,2}$,
taken from the experiments.

One can apply these methods to time-dependent flows.
Since the time-dependent circular jumps
typically involve breaking of the radial symmetry \cite{Ellegaard},
we take, as an example,
the two-dimensional (Cartesian) flow down an inclined plane.
There exists a large body of literature 
\cite{Kapitza,Benney,Pumir,Chang}
on such flows to which we shall be able to compare.
We non-dimensionalize in terms of 
the parabolic laminar solution \cite{Chang}
with a constant height $h_0$, mean velocity $v_0$, and flux $q_0$ 
related 
by $q_0 = h_0 v_0 = g h_0^3 \sin\alpha /(3 \nu)$, 
where $\alpha$ is the bottom slope.
The Reynold number is 
$R = v_0 h_0/\nu = q_0/\nu$ 
while the Froude number is 
$F = v_0^2/(g h_0 \cos\alpha) = R \tan\alpha /3$. 
We obtain \cite{long}
\begin{equation}
\label{td1}
  h_t + (hv)_x =0
\end{equation}
\begin{eqnarray}
  \frac{R}{3h} \left[ (hv)_t + ( h v^2 F_2(\lambda) )_x \right]
    + h_x \cot \alpha \nonumber \\
   = 1 - (\lambda+3) v/(3h^2)
  \label{td2}
\end{eqnarray}
\begin{equation}
\label{td3}
  h_x \cot \alpha = 1 - (5 \lambda+3)v/(3h^2)
\end{equation}
where $x$ is the scaled downward distance along the plane.

\begin{figure}[tbp]
\centerline{\psfig{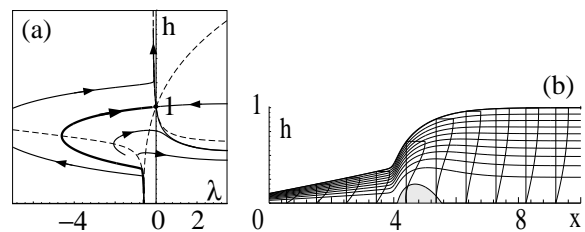}}
\vspace{3mm}
\caption{
Stationary solutions to the inclined plane equations
(\protect{\ref{td1}}--\protect{\ref{td3}})
with $R=30$ and $\alpha=1$[deg].
Phase portrait (a) has a saddle fixed point at 
$(h,\lambda)=(1,0)$.
Dashed curves are nullclines $h'=0$ or $\lambda'=0$.
Among trajectories (solid curves), one stable manifold
(drawn thicker)
to the saddle point corresponds to the hydraulic jump solution
shown in (b).
}
\label{cartesianjump}
\end{figure}

We first study stationary solutions to the equations.
Then, $hv=1$ from (\ref{td1}),
and (\ref{td2},\ref{td3}) form
an autonomous two-dimensional system for $(\lambda,h)$,
that can be easily studied on a phase portrait.
(This is the Cartesian version of (\ref{lambda1},\ref{lambda2}).)
There is a unique fixed point $h=1$ and $\lambda=0$,
and thus one cannot find stationary states 
connecting two different states with constant $(h,\lambda)$. 
(This can, however, be done for traveling waves;
see below.)
On the other hand an interesting solution \cite{Higuera}
is represented by the stable manifold of the fixed point 
emerging from $h=0$ as shown in Fig.~\ref{cartesianjump}.
The first part of the 
trajectory has $h_x$ nearly constant \cite{Watson},
i.e.\ $h \approx A (x-x_0)/R$ and $\lambda \approx -3/5$ 
until it suddenly jumps up to the fixed point values. 
Inserting into (\ref{td1},\ref{td2}), we get 
$A=2.4/F_2(-3/5) \approx 1.93$.
We believe this represents flows that 
are observed behind sluice gates
though we treat the flow as laminar \cite{turbulence}.
The conventional hydraulic theory predicts \cite{hydref} that
a jump occurs behind a gate when the bottom slope is ``mild'' 
(i.e.\ $\alpha$ being less than a critical slope).
Correspondingly, 
the jump structure in Fig.~\ref{cartesianjump}
disappears as $R \tan \alpha = 3F$ is increased 
beyond $A$ \cite{long}.

\begin{figure}[tbp]
\centerline{\psfig{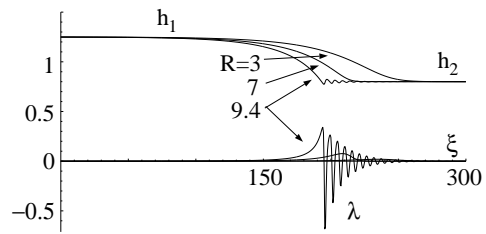}}
\vspace{3mm}
\caption{
Traveling wave solutions to the inclined plane equations.
$h_1=1/h_2=5/4$, $\alpha=2$[deg], and $R=3,7,9.4$.
The $u$-profile stays near parabolic ($\lambda=0$).
Oscillation starts when $R \tan \alpha$ is near
a critical value. 
}
\label{heteroclinic}
\end{figure}

It is also possible to find a traveling wave
solution \cite{Pumir}
which connects two parabolic laminar solutions 
of height $h_1$ at $x=-\infty$ and 
$h_2$ at $x=\infty$ \cite{homoclinic}. 
These two limits thus carry different fluxes, 
such that the flux is conserved in the moving frame only.
By choosing the characteristic height appropriately,
we may set $h_1 h_2 = 1$ without loss of generality.
Then we define the moving frame by $\xi = x - c t$,
and look for a stationary solutions in $\xi$, which by 
(\ref{td1}) must satisfy
$c = h_1^2 + h_1 h_2 + h_2^2 (> 3)$. 
There are two fixed points $h=h_{1,2}$, both with 
$\lambda=0$,
and a heteroclinic solution from $h_1$ to $h_2(<h_1)$ as $\xi$ increases
can be found \cite{long} iff
$R \tan \alpha < 60 h_1^3/( 25 c^2 h_1^4 - 61 c h_1^2 + 33 )$.
Such river-bore like solutions are calculated and shown
in Fig.~\ref{heteroclinic} for a fixed $h_1$ and $\alpha$
and varying $R$.
Note that the velocity profile always remains near parabolic
as $\lambda$ departs only slightly from zero,
and that the width of the ``shock'' 
is much larger than the thickness of the layer,
unlike the steady jump (Fig.~\ref{cartesianjump}).

Finally, we study the dispersion of small disturbances
in the time-dependent system.
The spectrum of the uniform
state ($h=v=1$, $\lambda=0$)
allows the distinction 
between super- and subcritical flows,
which is fundamental to hydraulics 
but not obvious for viscous flows.
Surface tension is necessary
for the stability calculations.
An additional term $(+ R W h_{xxx}/3)$ thus appears
on the right hand sides of (\ref{td2},\ref{td3}),
where $W = \sigma / (\rho h_0 v_0^2) = 
9 \sigma \nu^2/(\rho g h_0^5 \sin^2 \alpha)$ 
is the Weber number.

Assuming that all disturbances vary like $\exp (i kx - i \omega t)$,
we obtain \cite{long} two dispersion branches.
In the $k \rightarrow 0$ limit they behave as
$\omega_+(k) \sim 3 k + i k^2 (5 R/4 - \cot \alpha) + O(k^3)$ and 
$\omega_-(k) \sim -14k/25 - 12i/5R + O(k^3)$.
Thus, the $\omega_-$ branch moves backwards,
and the flow is, irrespective of the Froude number, 
``subcritical". 
From the imaginary parts, both branches are stable for a small $R$,
but the $\omega_+$ branch becomes unstable for $R \tan\alpha > 
4/5$.
This is in qualitative agreement with other models,
notably those coming from 
perturbation expansions \cite{Benney,Pumir,c3:oneway} 
and from averaging \cite{Chang}. 
The so-called ``Shkadov model" \cite{Chang} is identical 
to our system (\ref{td1},\ref{td2}) 
with a rigid parabolic profile, i.e.\ $\lambda \equiv 0$,
and omitting (\ref{td3}).

For very large $k$ the model shows unphysical behavior, 
as one branch becomes unstable.
{\em A priori} we have no reason 
to expect our model to be well defined at length scales much 
smaller than the normalized height, 
since our starting point is the boundary layer approximation. 
High-frequency oscillations are expected {\em not} to 
penetrate far into the fluid,
but our assumption of the hydrostatic pressure
still connects $\lambda$ and $h$ rigidly.
We can remedy this by modifying (\ref{td3}), 
such that $\lambda$ depends on a spatial average of $h$ and $h_x$ 
over an interval of the order of a fraction of $h$ \cite{c7}. 
In this way, the limit $k \rightarrow \infty$ now 
corresponds to the Shkadov model \cite{Chang} and is stable.

\begin{figure}[tbp]
\centerline{\psfig{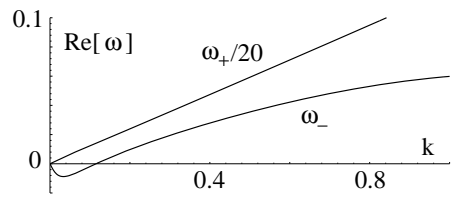}}
\vspace{3mm}
\caption{Real part of the dispersion relation $\omega(k)$.}
\label{dispersion}
\end{figure}

For intermediate $k$, the dispersion behavior is very interesting.
When $R \tan \alpha > 20/11 \approx 1.82$,
the group velocity of both branches 
will have positive real parts,
and the flow is ``supercritical''.
In terms of the Froude number,
this inequality becomes $F>20/33$
which is similar to the classical criterion of $F>1$
even though our theory includes viscosity.
For large $R \tan \alpha $, the subcritical range of $k$ 
is so small
that this supercritical behavior dominates.
In Fig.~\ref{dispersion},
we show dispersion curves (the real parts of $\omega_{\pm}$)
for $R=25$, $\alpha=5$~[deg], and $W=0.01$.
The slope of the $\omega_+$ branch is always positive.
On the other hand, the slope of the $\omega_-$ branch 
changes its sign.
The subcritical small-$k$ region is small,
and such long-wave disturbances become hard to create.
Note that the criterion $R \tan \alpha = 3F < A \approx 1.93$
for the existence of a stationary jump 
(Fig.~\ref{cartesianjump}) is almost equivalent to the demand
that the final state $h=1$ be subcritical
($R \tan \alpha < 1.82$). 
On the other hand,
the linearly increasing part before the jump 
in Fig.~\ref{cartesianjump} is expected to be supercritical,
but the dispersion around the solution is
hard to obtain due to its finite extent and the non-uniform
character.
In contrast, for the moving jumps in Fig.~\ref{heteroclinic},
super- or sub-criticality must be determined with respect
to the jump.
It can be shown \cite{long} that the flow is
supercritical in front of the jump
and subcritical behind, as expected.

To conclude, 
we have presented a simple model of free-surface flows
which can describe separation
and the structure of the circular and
linear hydraulic jumps.

We thank Clive Ellegaard, Adam E.\ Hansen and Anders Haaning 
for many discussions and for providing us with
experimental data. 
We would also like to thank Ken H.\ Andersen,
Peter Dimon, Lisbeth Kjeldgaard 
and Hiraku Nishimori for stimulating discussions.

\begin{references}
\bibitem{c1:riverbore} 
Similar moving structures are 
usually called river bores.
\bibitem{turbulence} 
It is presumably possible to replace the loss term
in (\ref{td1}-\ref{td3}) to model turbulent flows, as e.g.\ in 
\cite{Whitham}.
\bibitem{homoclinic} 
It is also possible to study homoclinic orbits \cite{Pumir,Chang}, 
but there surface tension appears crucial.
\bibitem{c3:oneway} 
The model used in \cite{Pumir} has only one 
branch of dispersion relation
and thus no distinction between  supercritical and
subcritical flow.
\bibitem{c7} 
The modified equation (\ref{td3}) is 
$1 - (5 \lambda+3)v/3h^2 = \int_{-\infty}^{+\infty} g(x-s) p(s,t) ds$
where $p(x,t) = h_x \cot \alpha - RW h_{xxx}/3$
and $g(x)= g_0 \exp (-\pi g_0^2 x^2)$. 
In the $g_0 \rightarrow \infty$ limit 
this reduces to (\ref{td3}) and
$k_c = 2 \sqrt{\pi} g_0$ signifies the cutoff in $k$, which, in 
Fig.~\ref{dispersion} was $k_c=6\pi$.
\end{references}
\end{document}